\documentclass[reprint,amsmath,amssymb,aps]{revtex4-2}

\usepackage[hidelinks]{hyperref}
\usepackage{natbib}
\usepackage{graphicx}
\usepackage{dcolumn}
\usepackage{bm}
\usepackage{textcase}
\usepackage{url}
\usepackage[mathlines]{lineno}
\newcolumntype{C}[1]{>{\centering\arraybackslash}p{#1}}

\begin{document}

\preprint{APS/123-QED}

\title{Energy dependence of particle production in Au + Au collisions \\ at $\sqrt{s_{\text{NN}}}$ = 7.7 - 200 GeV using a multiphase transport model}

\author{Krishan Gopal}
\email{krishangopal@students.iisertirupati.ac.in}
\author{Chitrasen Jena}
\email{cjena@iisertirupati.ac.in}
\author{Kishora Nayak}
\email{k.nayak1234@gmail.com}
\thanks{\\Presently at Department of Physics, Panchayat College, Sambalpur University, Bargarh 768028, Odisha, India}
\affiliation{Department of Physics, Indian Institute of Science Education and Research (IISER) Tirupati, Tirupati 517507, Andhra Pradesh, India}

\date{\today}

\begin{abstract}
In this study, we employ a multi-phase transport (AMPT) model to understand the production of $\pi^{\pm}$, $K^{\pm}$, $p$, $\overline{p}$, $K^{0}_{s}$, $\Lambda$, $\bar{\Lambda}$, and $\phi$ in Au + Au collisions at $\sqrt{s_{NN}} = 7.7$, $27$, $39$, $62.4$, and $200$ GeV. We have studied the energy dependence of various bulk properties of the system such as transverse momentum ($p_T$) spectra, particle yields ($dN/dy$), mean transverse mass ($\langle m_T \rangle$), and anti-particle to particle ratios. Model calculations using both default and string melting versions of the AMPT with three distinct sets of initial conditions are compared to the data from the STAR experiment. In the case of $\pi^{\pm}$, $K^{\pm}$, $p$, and $\overline{p}$, we observe that the string melting version shows better agreement with data at higher energies, while the default version performs better at lower collision energies. However, for $K^{0}_{s}$, $\Lambda$, and $\phi$, it is observed that the default version is able to describe the data better at all energies. In addition, we have used the blast-wave model to extract the kinetic freeze-out properties, like the kinetic freeze-out temperature and the radial flow velocity. We observe that these parameters are comparable with the data.

\begin{description}
\item[keywords]
AMPT Model,  particle spectra,  bulk properties, kinetic freeze-out properties.
\end{description}
\end{abstract}

\maketitle


\section{\label{sec:level1}Introduction}

Heavy ion collision experiments provide an opportunity to probe the behaviour of nuclear matter under extreme conditions of energy density and temperature. Quantum Chromodynamics (QCD), the theory governing the strong nuclear force, predicts that under the extreme energy densities achieved in heavy-ion collisions, the hadronic matter transitions to a deconfined state of quarks and gluons, known as Quark-Gluon Plasma (QGP) \cite{SHURYAK198071,PHENIX:2004vcz,STAR:2005gfr,BRAHMS:2004adc,BACK200528}. Experimental facilities such as the Relativistic Heavy Ion Collider (RHIC) at the Brookhaven National Laboratory (BNL) and the Large Hadron Collider (LHC) at the European Organization for Nuclear Research (CERN), aim to study various properties of the QGP. 
One central goal of these experiments is to map out the QCD phase diagram, typically 
plotted as temperature ($T$) versus baryon chemical potential ($\mu_B$). In that regard, RHIC initiated the Beam Energy Scan (BES) program in 2010, collecting the data in Au+Au collisions at $\sqrt{s_{NN}}$ = 7.7 -- 200 GeV. BES program spanned a wide range of $\mu_B$ from 20 to 420 MeV \cite{PhysRevC.86.054908, PhysRevC.88.014902, PhysRevLett.110.142301, PhysRevLett.112.162301, PhysRevLett.113.052302, PhysRevC.93.021903, PhysRevC.96.044904, PhysRevC.96.064905,PhysRevC.93.024901, PhysRevC.93.011901, PhysRevLett.109.152301}. Lattice QCD calculations predict that the phase transition from the hadronic matter to QGP occurs somewhere in this range of $\mu_B$ \cite{PhysRevD.78.074507}. 

In this study, the bulk properties of the system in
Au+Au collisions at $\sqrt{s_{NN}}$ = 7.7, 27, 39, 62.4, and 200 GeV are studied using a multi-phase transport (AMPT) model. We employ three distinct sets of input parameters for both the string melting (AMPT-SM) and the default (AMPT-Def) versions of the AMPT {\cite{PhysRevC.72.064901}}. Bulk properties, such as transverse momentum ($p_{T}$) spectra, integrated yields ($dN/dy$), mean transverse mass ($\langle m_T \rangle$), and particle ratios at various collision energies are studied. We also study the kinetic freeze-out (KFO) parameters obtained by fitting the transverse momentum spectra using a hydrodynamically inspired blast-wave model {\cite{PhysRevC.48.2462}}. KFO refers to the surface of last scattering of hadrons produced in heavy-ion collisions, effectively defining the surface of their final momentum distribution. Freeze-out properties provide information on the evolution of the medium produced in heavy-ion collisions.

The paper is organised as follows: Section~\ref{sec:level2} provides a brief overview of the AMPT model. In Sec.~\ref{sec:level3}, we discuss the comparison of AMPT model results with the experimental data. Section~\ref{sec:A} compares the $p_T$ spectra of various particles obtained from the AMPT model with experimental data. Section~\ref{sec:B} presents the energy dependence of $dN/dy$, $\langle m_T \rangle$, and particle ratios at various collision energies. In Sec.~\ref{sec:C}, we show the dependence of freeze-out parameters on the collision centrality and energy. Finally, Section~\ref{sec:level4} summarises our findings.  

\section{\label{sec:level2}AMPT Model}
AMPT is a hybrid monte carlo event generator that allows the study of the dynamics of relativistic heavy-ion collisions. It has been widely used to analyse various collision systems at different centre-of-mass energies 
\cite{Ye:2017ooc, PhysRevC.102.024902, TIWARI2020121751}. AMPT model generates fluctuating initial conditions using the Heavy Ion Jet INteraction Generator (HIJING) model, which involves spatial and momentum distributions of minijet partons and soft string excitations \cite{PhysRevD.44.3501}. Evolution of partonic interactions is modelled using the Zhang’s Parton Cascade (ZPC) \cite{Zhang:1997ej}. The differential cross-section of parton-parton scattering can be expressed as: 
\begin{equation}
  \frac{d\sigma}{dt} \approx \frac{9\pi\alpha_s^2}{2(t-\mu^2)^2},
\end{equation}
where $\sigma$ is cross section of partonic two body scattering, $t$ is the standard Mandelstam variable for four-momentum transfer, $\alpha_s$ is the strong coupling constant, and $\mu$ is the Debye screening mass in the partonic matter. The \(\mu\) is influenced by the medium effects, in turn affecting $\sigma$. A medium dependent \(\mu\) would produce results that are closer to the experimental data. In addition, it has been observed that particle multiplicity in a collision is not very sensitive to \(\sigma\), but it might affect the elliptic flow such that a larger $\sigma$ leads to larger elliptic flow \cite{Xu:2011fi}.


The two versions of AMPT, Default and String Melting, differ in how they treat the formation of hadrons. In the AMPT-Def version, the formation of hadrons from quarks and antiquarks is governed by Lund String model employing a symmetric fragmentation function:
\begin{equation}
  f(z) \propto z^{-1}(1-z)^a \exp(-bm^2_T / z),
\end{equation}
where $z$ is the light-cone momentum fraction of the produced hadrons relative to the fragmenting string, while $a$ and $b$ are called Lund string fragmentation parameters \cite{Andersson:1983ia}. The average squared transverse momentum, $\langle p^2_T \rangle$, of the produced particles is proportional to the string tension $\kappa$, which represents the energy stored per unit length of a string:
\begin{equation}
 \langle p_T^2 \rangle \propto \kappa = \frac{1}{b(2 + a)}.
\end{equation}
According to Eq. (3), the parameters  \(a\) and \(b\) determine the \(p_{T}\) distribution of particles produced in heavy-ion collisions. A large value of \(a\) and/or \(b\) will lead to a small \(\langle p_{T}^{2} \rangle\), resulting into a sharp \(p_{T}\) spectrum. However, a smaller value of \(a\) and \(b\) will produce a flat \(p_{T}\) distribution \cite{PhysRevC.102.024902}.

The AMPT-SM version, on the other hand, is based on the concept that beyond a certain critical energy density ($\sim$ 1 GeV/c), the coexistence of colour strings and partons becomes unfavourable. Consequently, these strings “melt” into low momentum partons at the start of ZPC. The transported partons finally undergo hadronization by spatial quark coalescence mechanism \cite{PhysRevC.96.014910}.

 \begin{table}[ht]
  \centering
  \begin{tabular}{C{0.91cm}C{0.91cm}C{1.5cm}C{1.1cm}C{1.25cm}C{1.5cm}}
    \hline \hline
     & $a$ & $b$($GeV^{-2}$) & $\alpha_s$ & $\mu (fm^{-1})$ & $\sigma$ (mb) \\ \hline
    Set 1 & 0.55 & 0.15 & 0.33 & 2.265 & 3\\ \hline
    Set 2  & 2.2 & 0.15 & 0.33 & 2.265 & 3 \\ \hline
    Set 3 & 0.5 & 0.9 & 0.33 & 3.2 & 1.5 \\ \hline \hline
  \end{tabular}
  \caption{Different sets of input parameters for Lund string fragmentation and parton scattering cross sections used in this study.}
  \label{table:5x5}
\end{table}

The subsequent interaction of hadronic matter is characterised by a hadronic cascade, which is governed by a relativistic transport (ART) model \cite{PhysRevC.52.2037}. 

Table~\ref{table:5x5} lists various choices of input parameters of the AMPT model we have used in this work to study particle production in heavy-ion collisions. The different choices of these parameters were made by insights derived from previous studies \cite{Ye:2017ooc, PhysRevC.102.024902, TIWARI2020121751}. We have performed a systematic study of particle $p_T$ spectra using various set of input parameters as mentioned below, 
\begin{itemize}
    \item Set-1 and Set-2 differ in the \(a\) parameter while keeping \(b\) and \(\sigma\) are same. A larger value of \(a\) is expected to yield a sharp \(p_T\) spectrum.
    \item In Set-1 and Set-3, all three parameters \(a\), \(b\), and \(\sigma\) are different. We have decreased all these three values to observe their effect on the bulk properties.
\end{itemize}
We aim to find the best set of input parameters that properly describe the experimentally measured \(p_{T}\)  spectra. In the following section experimental results are compared with the results from different configurations of the AMPT model.

\section{\label{sec:level3}Results}
We report the mid-rapidity \(p_T\) spectra, \(dN/dy\), \(\langle m_T \rangle\) of $\pi^{\pm}$, $K^{\pm}$, $p$, $\overline{p}$, $K^{0}_{S}$, $\Lambda$, and $\phi$, along with particle ratios in most central Au+Au collisions at \(\sqrt{s_{NN}} = 7.7\), \(27\), 39, 62.4, and \(200\) GeV. The results are obtained for both AMPT-SM and AMPT-Def configurations. We compare the model results to those obtained from the STAR experiment.




\subsection{\label{sec:A} Transverse momentum spectra}

\begin{figure*}
  \centering
  \includegraphics[height=9cm,width=0.9\textwidth]{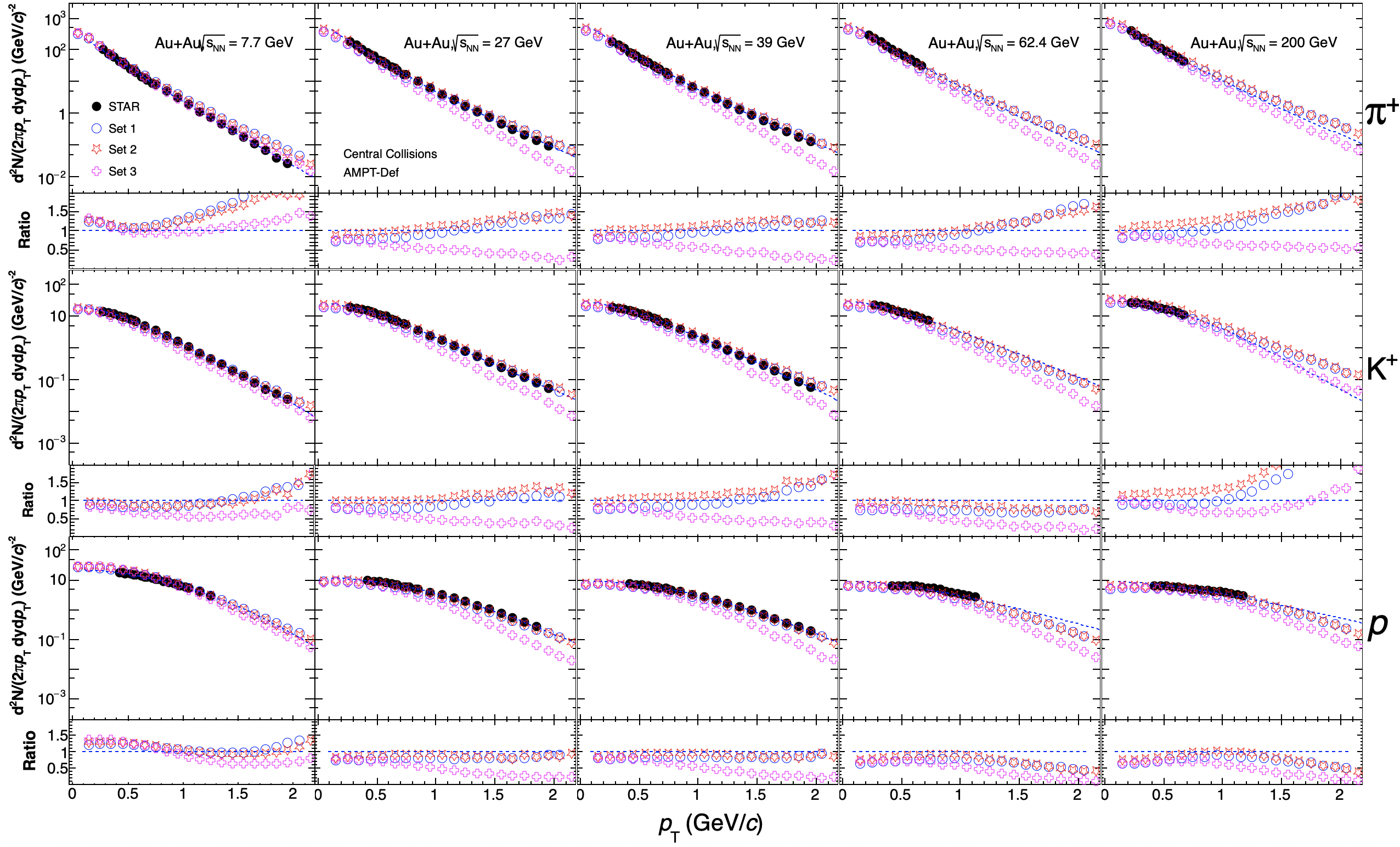}
  \caption{Mid-rapidity ($|y|<0.1$) invariant yield of $\pi^{+}$ (upper row), $K^{+}$ (middle row) and proton (lower row) as a function of $p_{T}$ for  $0$-$5$\% centrality  in Au+Au collision at $\sqrt{s_{\text{NN}}}$ = 7.7, 27, 39, 62.4, and 200 GeV using AMPT-Def version. The AMPT results are compared with the corresponding experimental data which is fitted with the Levy-Tsallis function~\cite{PhysRevC.102.034909}. The ratio of data to fit function is shown in the lower panels of each pad.}
  \label{fig:SpecRatio_piKp_AMPT_Default}
\end{figure*}

Figure~\ref{fig:SpecRatio_piKp_AMPT_Default} shows the comparison of mid-rapidity \(p_T\) spectra of $\pi^{+}$, $K^{+}$, and $p$ in most central Au+Au collisions obtained using the AMPT-Def with the STAR data at $\sqrt{s_{\text{NN}}} = 7.7$, 27, 39, 62.4, and 200 GeV \cite{PhysRevC.96.044904}. 
The experimental data is fitted with the Levy-Tsallis function and the lower panel in each plot shows the ratio of the invariant yield ($\frac{d^2N}{2\pi p_Tdydp_T}$) obtained from the fit function to the one obtained from AMPT-Def using different sets of input parameters. It is observed that both Set-1 and Set-2 appear to adequately describe the data, particularly at low energies, while Set-3 consistently underpredicts the data. 

\begin{figure*}
  \centering
  \includegraphics[height=9cm,width=0.9\textwidth]{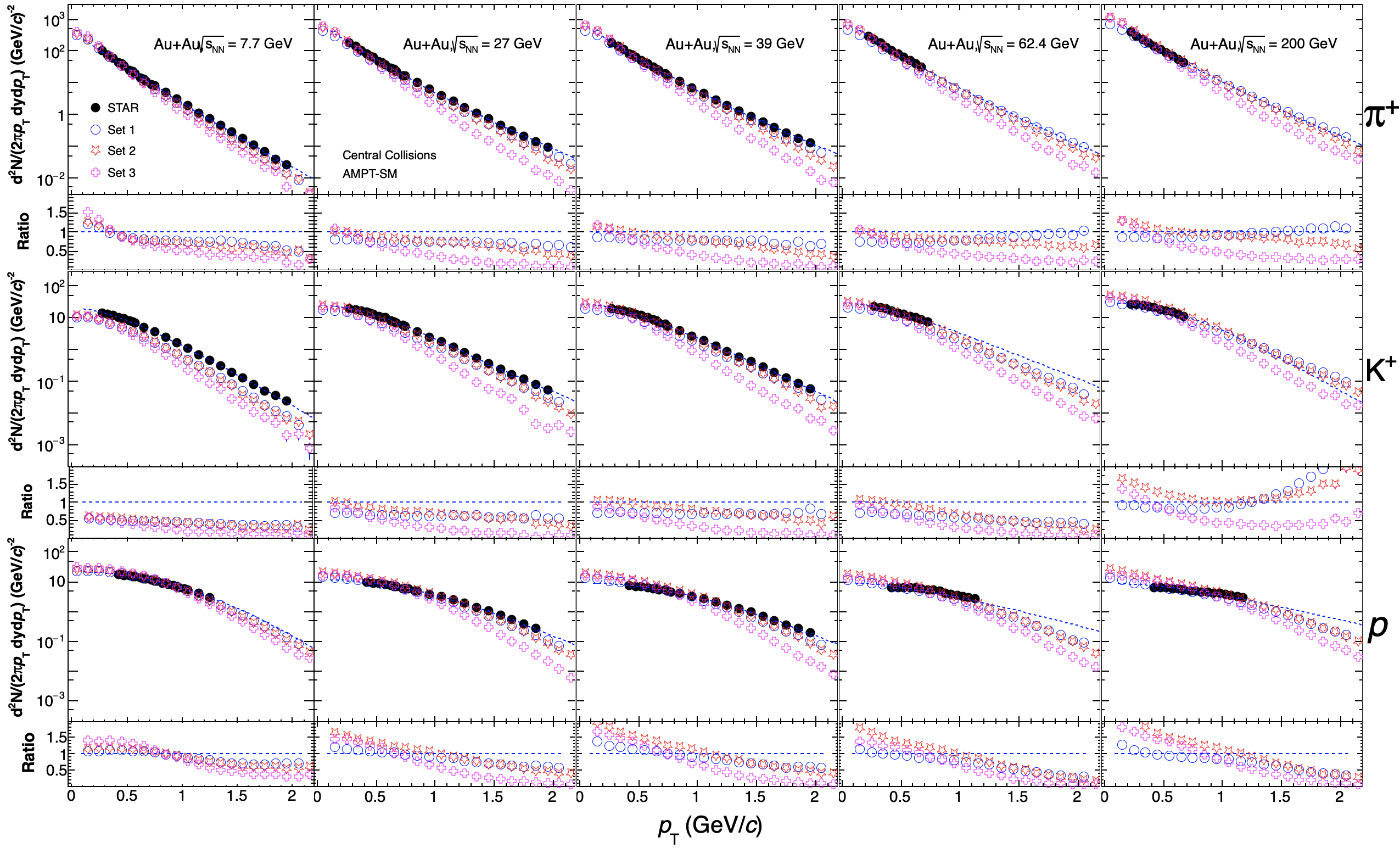}
  \caption{Mid-rapidity ($|y|<0.1$) invariant yield of $\pi^{+}$ (upper row), $K^{+}$ (middle row) and proton (lower row) as a function of $p_{T}$ for  $0$-$5$\% centrality in Au+Au collision at $\sqrt{s_{\text{NN}}}$ = 7.7, 27, 39, 62.4, and 200 GeV using AMPT-SM version. The AMPT results are compared with the corresponding experimental data which is fitted with the Levy-Tsallis function~\cite{PhysRevLett.84.2770}. The ratio of data to fit function is also shown in the lower panels for each pad.}
  \label{fig:SpecRatio_piKp_AMPT_SM}
\end{figure*}

Similar to Fig.~\ref{fig:SpecRatio_piKp_AMPT_Default}, Fig.~\ref{fig:SpecRatio_piKp_AMPT_SM} presents a comparison of mid-rapidity $p_T$ spectra of $\pi^{+}$, $K^{+}$, and $p$ in most central Au+Au collisions obtained using the AMPT-SM with the STAR data at $\sqrt{s_{\text{NN}}} = 7.7$, 27, 39, 62.4, and 200 GeV.
STAR data is fitted with Levy-Tsallis function and the lower panel in each plot shows the ratio of the invariant yield obtained from the fit function to the one obtained with AMPT-SM using different sets of input parameters. 
We observe that Set-2 provides an accurate description of the data at higher energies among all the three sets, however, it fails to describe the data at lower energies. 
We also observe that compared to $\pi^{+}$ and $p$, $p_T$ spectra of $K^{+}$ deviates significantly from the experimental data for all the different sets of AMPT-SM model considered in this study, perhaps due to the difference of strangeness quantum number. We have also explicitly checked that these result hold true for antiparticles as well. 

We also observe that the invariant yield shows a monotonically decreasing trend with increasing $p_{T}$ across all particles and sets of parameters for both AMPT-Def and AMPT-SM. In addition, the inverse slopes of particle spectra display a consistent trend: $\pi < K < p$ similar to what has been observed in experimental data.  

\begin{figure*}
  \centering  \includegraphics[height=10cm,width=0.9\textwidth]{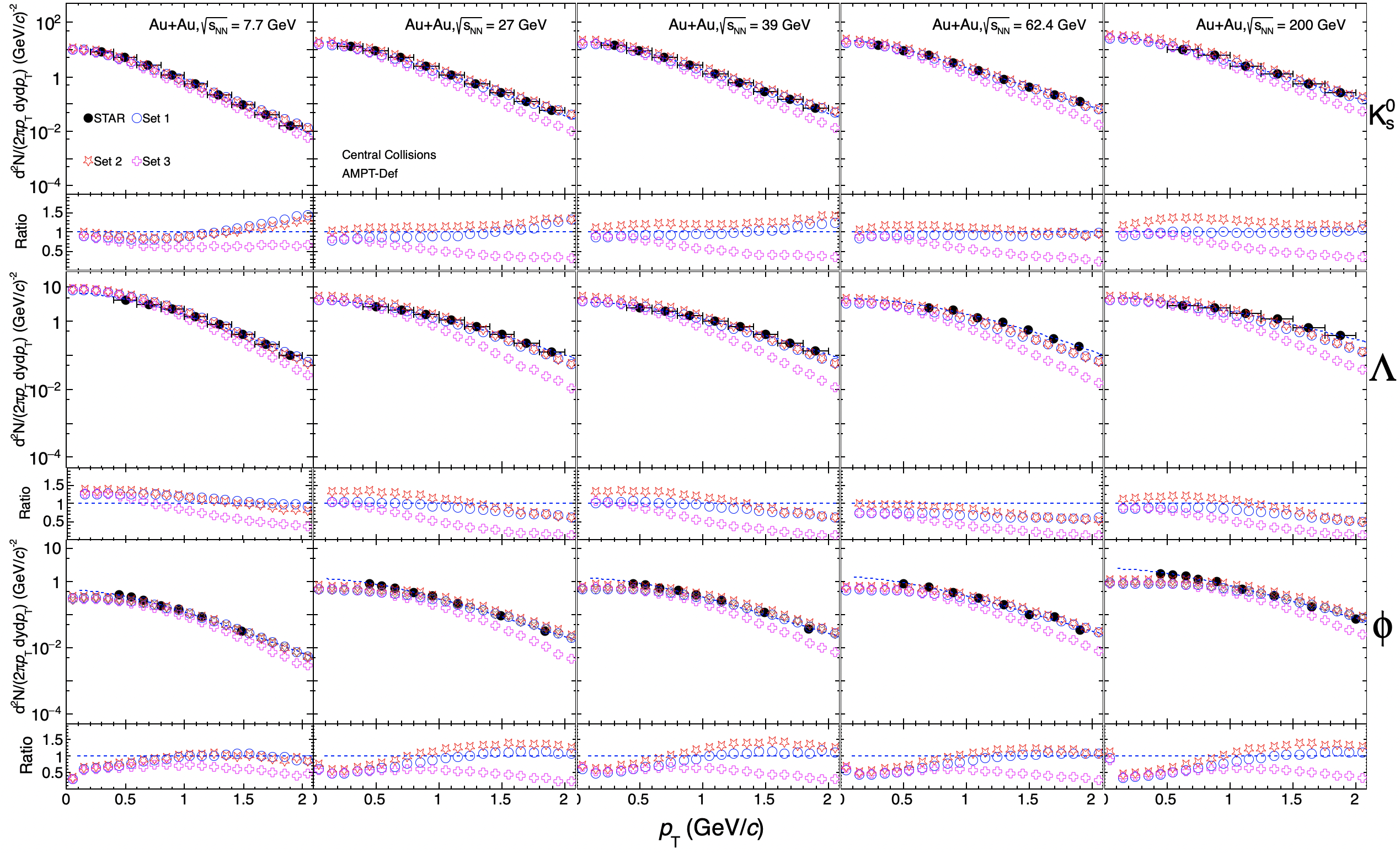}
  \caption{Invariant yield of $K^{0}_{S}$ (upper row), $\Lambda$ (middle row) and $\phi$ (lower row) as a function of $p_{T}$ for $0$-$5$\% centrality\footnote{$\phi$-meson spectra is calculated in $0$-$10$\% centrality of Au+Au collisions at $\sqrt{s_{NN}}$ = 7.7, 27, 29, and 200 GeV and $0$-$20$\% centrality at $\sqrt{s_{NN}}$ = 62.4 GeV as per the availability of the experimental data.} in Au+Au collision at $\sqrt{s_{\text{NN}}}$ = 7.7, 27, 39, 62.4, and 200 GeV at $|y|<0.5$ ($<1.0$, for $\Lambda$ in Au+Au collisions at $\sqrt{s_{NN}}$ = 200 GeV)) are shown in each column using AMPT-Def. The ratio of data to fit function is also shown in the lower panels for each pad.}
\label{fig:SpecRatio_K0s_Lam_Phi_AMPT_Deft}
\end{figure*}

\begin{figure*} 
  \centering  \includegraphics[height=10cm,width=0.9\textwidth]{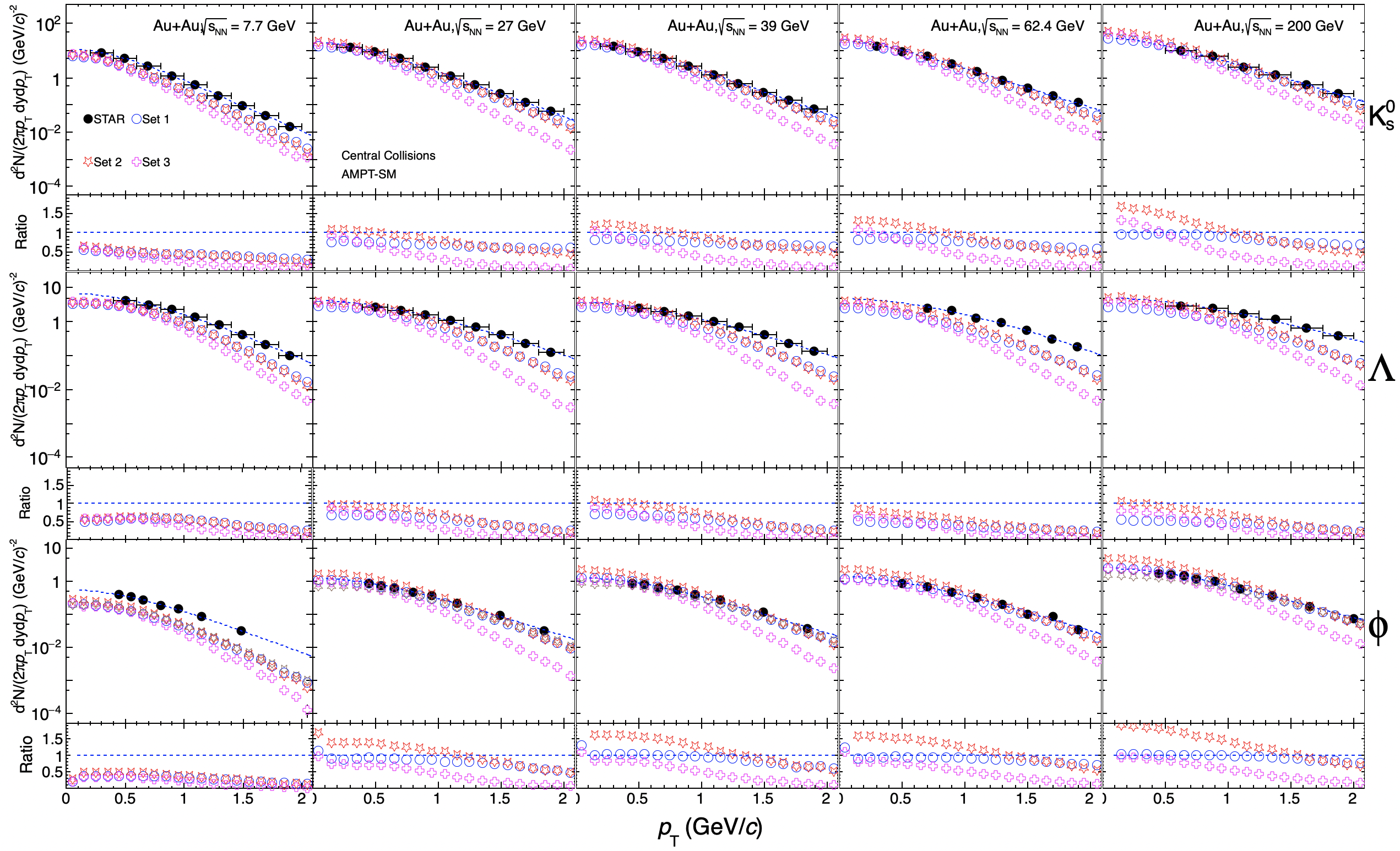}
  \caption{Invariant yield of $K^{0}_{S}$ (upper row), $\Lambda$ (middle row) and $\phi$ (lower row) as a function of $p_{T}$ for $0$-$5$\% centrality\footnote{$\phi$-meson spectra is calculated in $0$-$10$\% centrality of Au+Au collisions at $\sqrt{s_{NN}}$ = 7.7, 27, 29, and 200 GeV and $0$-$20$\% centrality at $\sqrt{s_{NN}}$ = 62.4 GeV as per the availability of the experimental data.} in Au+Au collision at $\sqrt{s_{\text{NN}}}$ = 7.7, 27, 39, 62.4, and 200 GeV at $|y|<0.5$ ($<1.0$, for $\Lambda$ in Au+Au collisions at $\sqrt{s_{NN}}$ = 200 GeV) are shown in each column using AMPT-SM. The ratio of data to fit function is also shown in the lower panels for each pad.}
  \label{fig:SpecRatio_K0s_Lam_Phi_AMPT_SM}
\end{figure*}

Figures \ref{fig:SpecRatio_K0s_Lam_Phi_AMPT_Deft} and \ref{fig:SpecRatio_K0s_Lam_Phi_AMPT_SM} 
show a comparison of mid-rapidity $p_T$ spectra of $K^{0}_s$, $\Lambda$, and $\phi$ obtained from the AMPT model with the STAR data in most central Au+Au collisions at $\sqrt{s_{\text{NN}}}$  = 7.7, 27, 39, 62.4, and 200 GeV \cite{PhysRevC.102.034909}. The experimental data is compared to predictions from both the AMPT-Def and AMPT-SM model with distinct sets of input parameters. STAR data is fitted with the Levy-Tsallis function and the lower panel in each plot shows the ratio of the invariant yield obtained from the fit function to the one obtained from the AMPT model. We observe that among the parameter sets, Set-1 in AMPT-Def effectively describes the $p_T$ spectra of strange hadrons. Whereas, in the case of AMPT-SM, no specific set is describing the data consistently.

\subsection{\label{sec:B} $p_{T}$ integrated yield, mean transverse mass, and particle ratios} 


Figure 5 shows a comparison of the energy dependence of $dN/dy$ of $\pi^{\pm}$, $K^{\pm}$, $p$, and $\bar{p}$, normalised by half of the average number of participating nucleons  ($\langle N_{\textrm{part}} \rangle / 2$) obtained using the AMPT model in the most central Au+Au collisions at $\sqrt{s_{\text{NN}}}$ = 7.7, 27, 39, 62.4, and 200 GeV with the STAR data. We observe that of all the choices of input parameters, Set-2 describes the particle yields reasonably well for both AMPT-Def and AMPT-SM versions as shown in Fig. 5(a) and Fig. 5(b), respectively. $dN/dy$ of $\pi^{\pm}$, $K^{\pm}$, $p$ and $\bar{p}$ is observed to increase with increasing energy. However, $dN/dy$ of $p$ decreases with increasing energy due to the baryon stopping prominently observed at lower energies.

\begin{figure*} 
  \centering
  \begin{minipage}[b]{0.48\textwidth}
    \includegraphics[height=7.5cm,width=\textwidth]{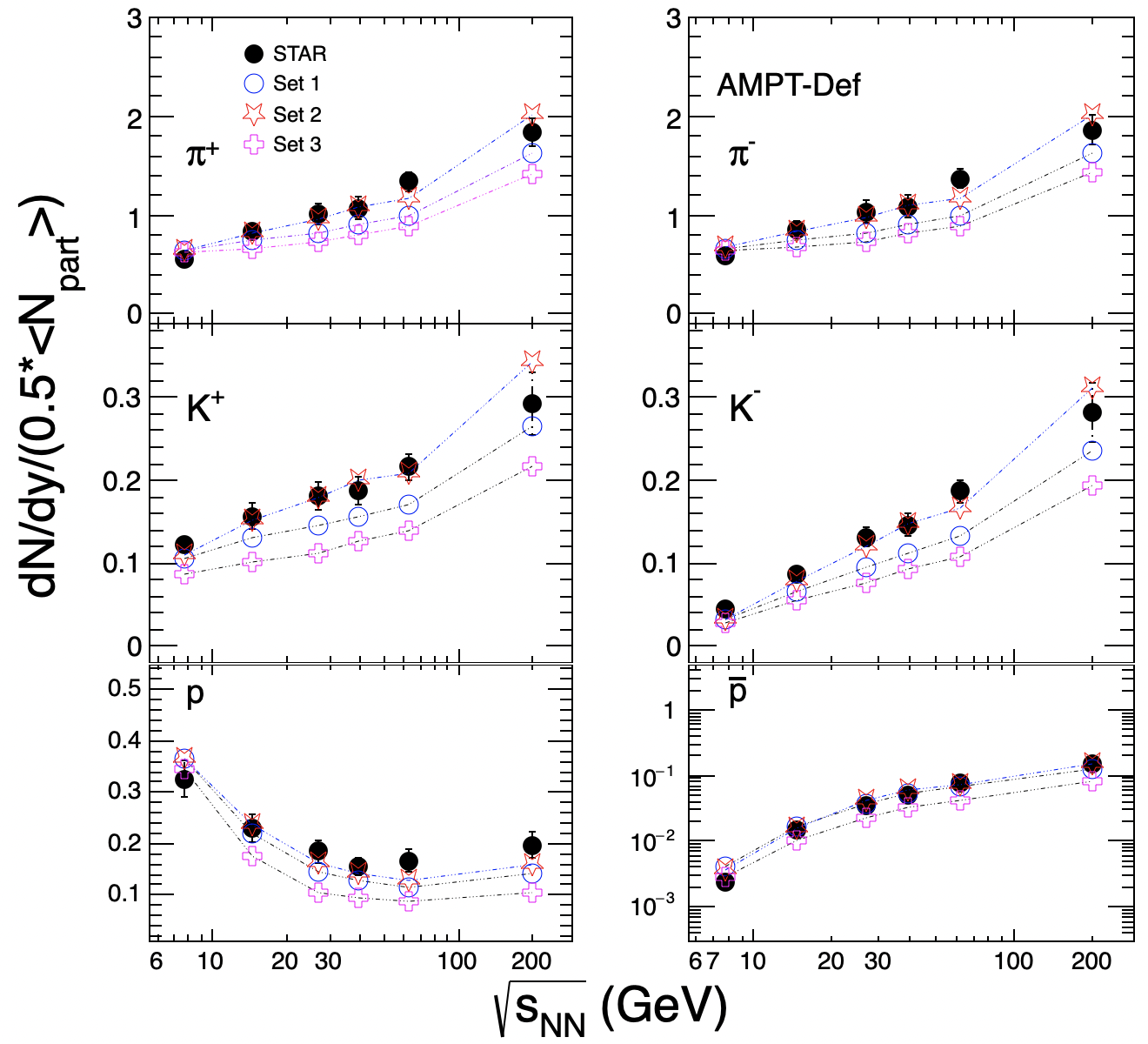}
    \text{(a)}
  \end{minipage}
  \hfill
  \begin{minipage}[b]{0.48\textwidth}
    \includegraphics[height=7.5cm,width=\textwidth]{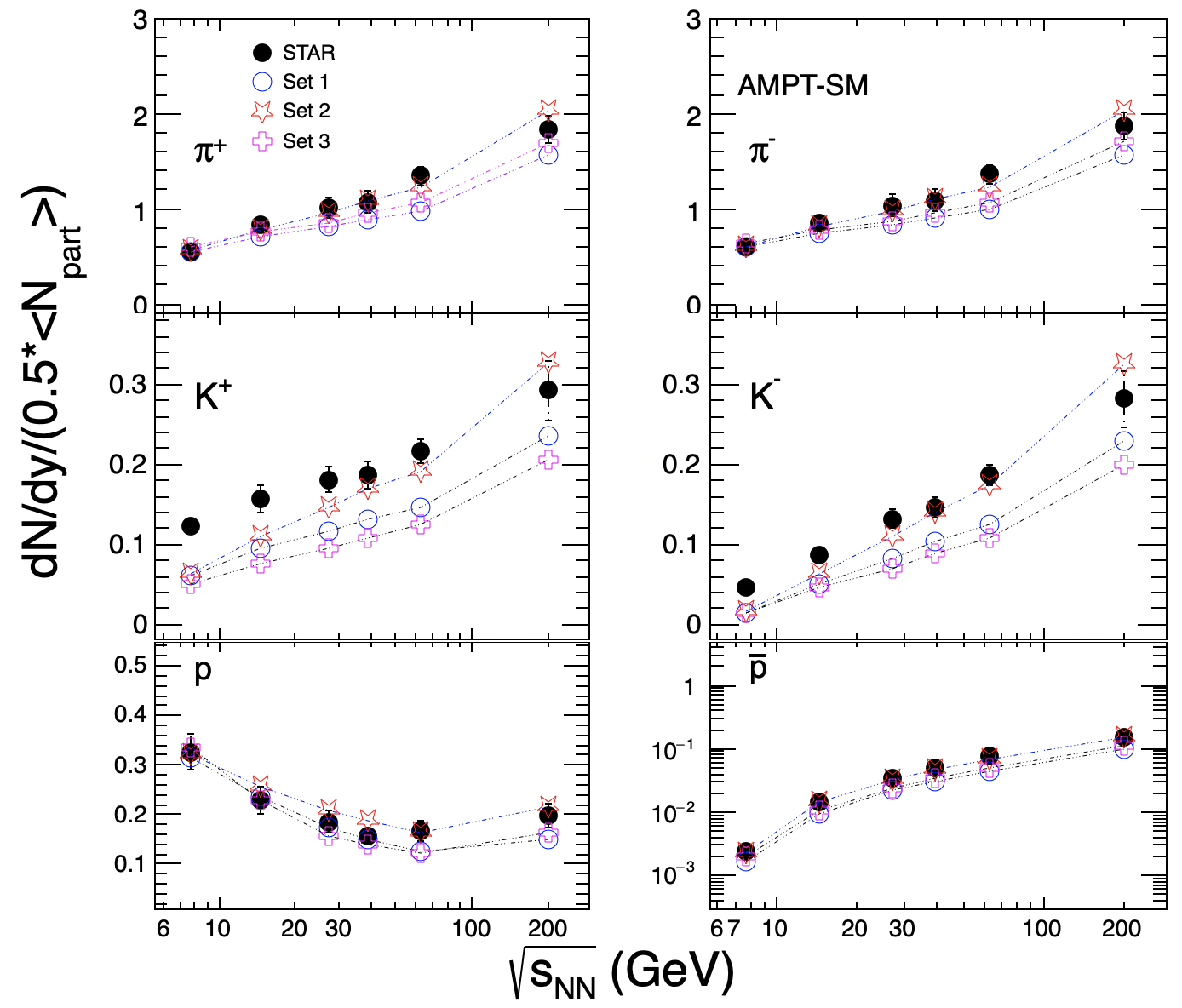}
    \text{(b)}
  \end{minipage}
  \caption{Energy dependence of (d$N$/dy)/($\langle N_{\textrm{part}} \rangle / 2$) for particle ($\pi^{+}$, $K^{+}$, $p$) and anti-particle ($\pi^{-}$, $K^{-}$, $\bar{p}$) for central Au+Au collisions using AMPT-Def (left panel) and  AMPT-SM (right panel). The AMPT results shown in open markers are compared with the corresponding experimental data (solid marker).}
  \end{figure*}

Figure 6 shows a comparison of the energy dependence of $dN/dy$ of $K_s^0$, $\Lambda$, $\bar{\Lambda}$, and $\phi$ normalised by \(\langle N_{\textrm{part}} \rangle / 2\) in the most central Au+Au collisions at $\sqrt{s_{\text{NN}}}$ = 7.7, 27, 39, 62.4, and 200 GeV calculated using AMPT model with the STAR data. We observe that Set-1 and Set-2 of AMPT-Def better describe the data for strange hadrons. 

\begin{figure*} 
  \centering
  \begin{minipage}[b]{0.48\textwidth}
    \includegraphics[height=6cm,width=\textwidth]{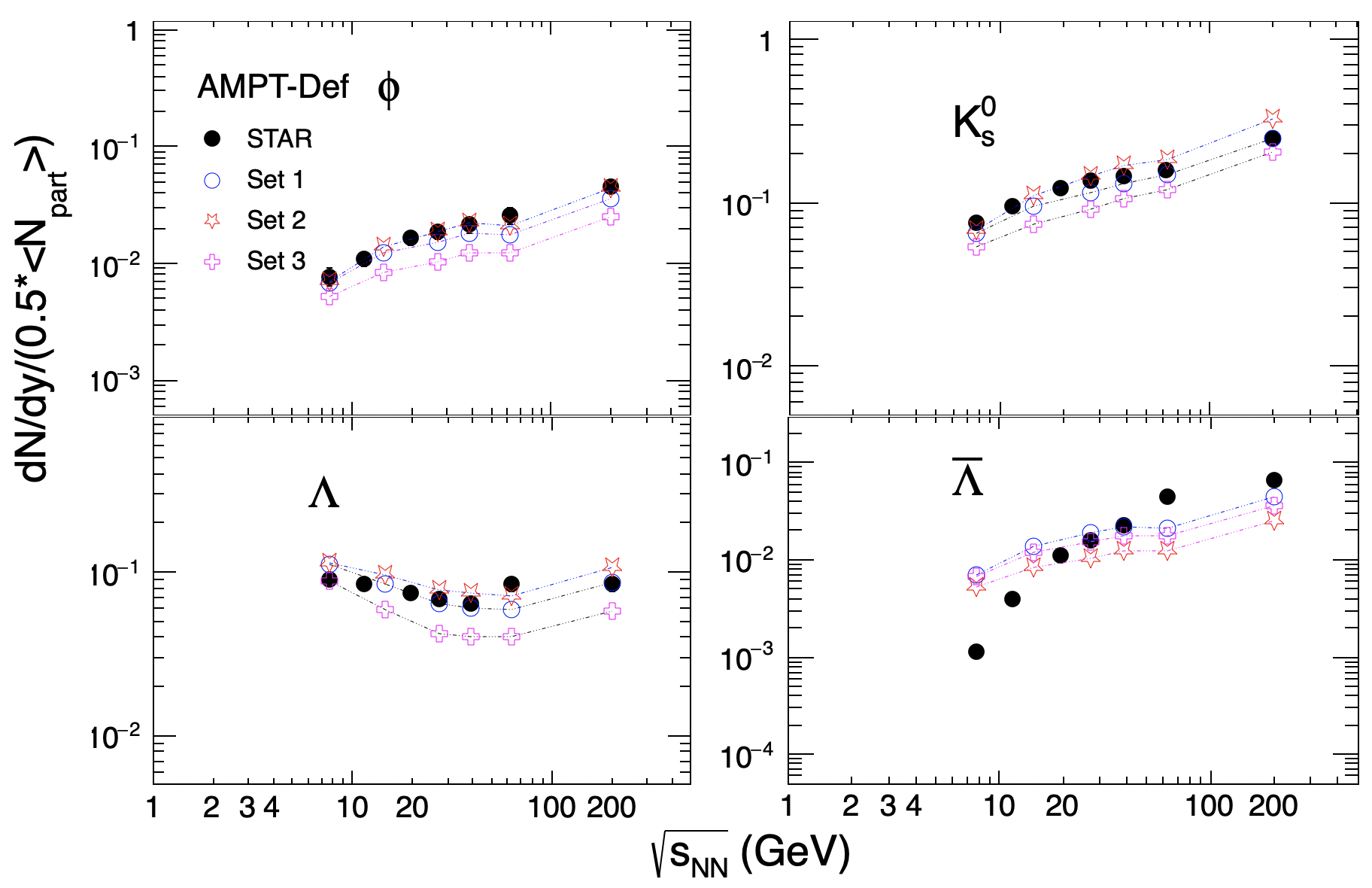}
    \text{(a)}
  \end{minipage}
  \hfill
  \begin{minipage}[b]{0.48\textwidth}
    \includegraphics[height=6cm,width=\textwidth]{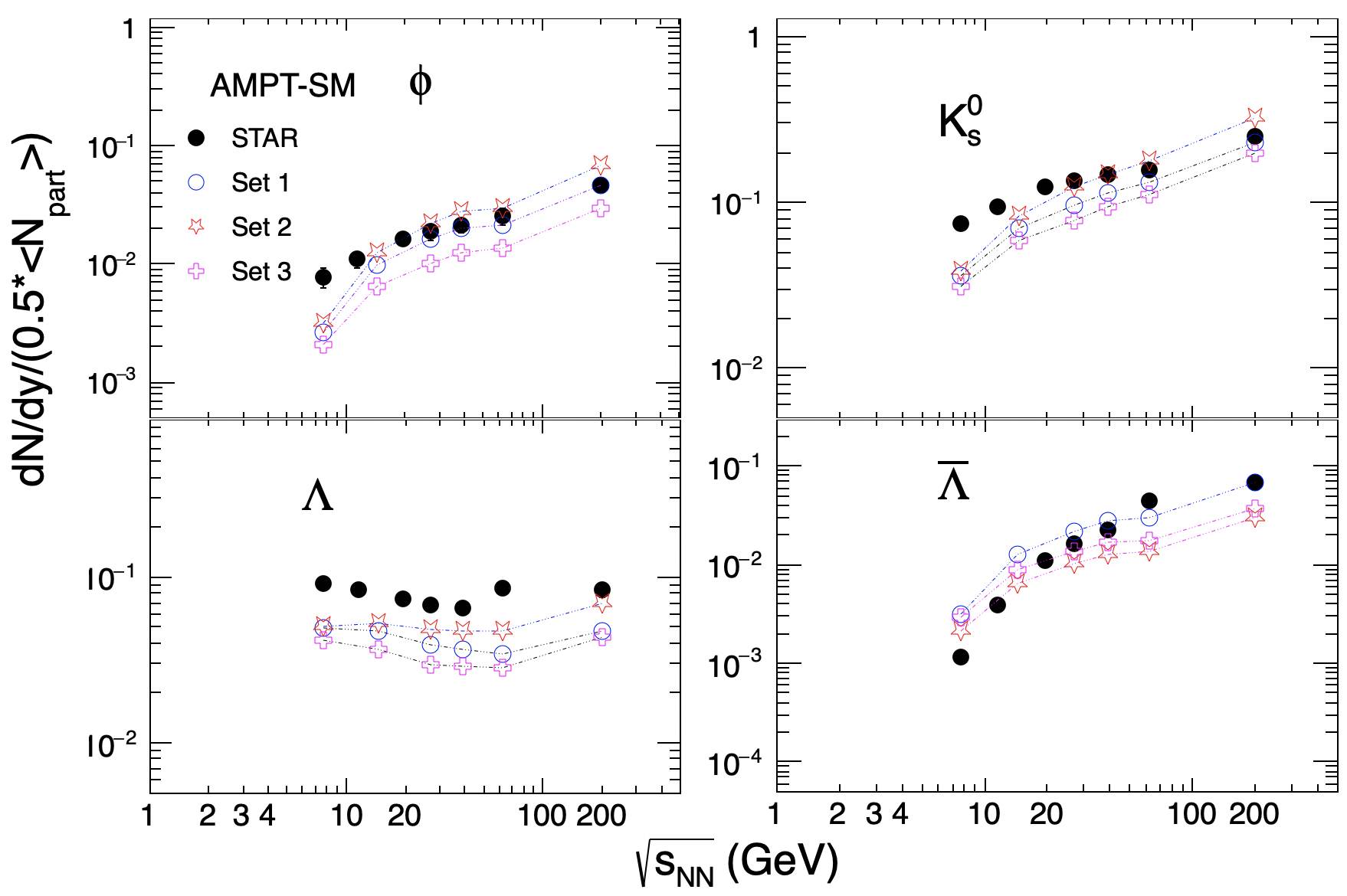}
    \text{(b)}
  \end{minipage}
  \caption{Energy dependence of (d$N$/dy)/($\langle N_{\textrm{part}} \rangle / 2$) for  $K^{0}_{S}$, $\Lambda$, $\phi$, and $\bar{\Lambda}$
 for central Au+Au collisions using AMPT-Def (left panel) and  AMPT-SM (right panel). The AMPT results shown in open markers are compared with the corresponding experimental data (solid marker).}
  \end{figure*}

\begin{figure*}[!hbtp]
  \centering
  \begin{minipage}[b]{0.48\textwidth}
    \includegraphics[height=7.5cm,width=\textwidth]{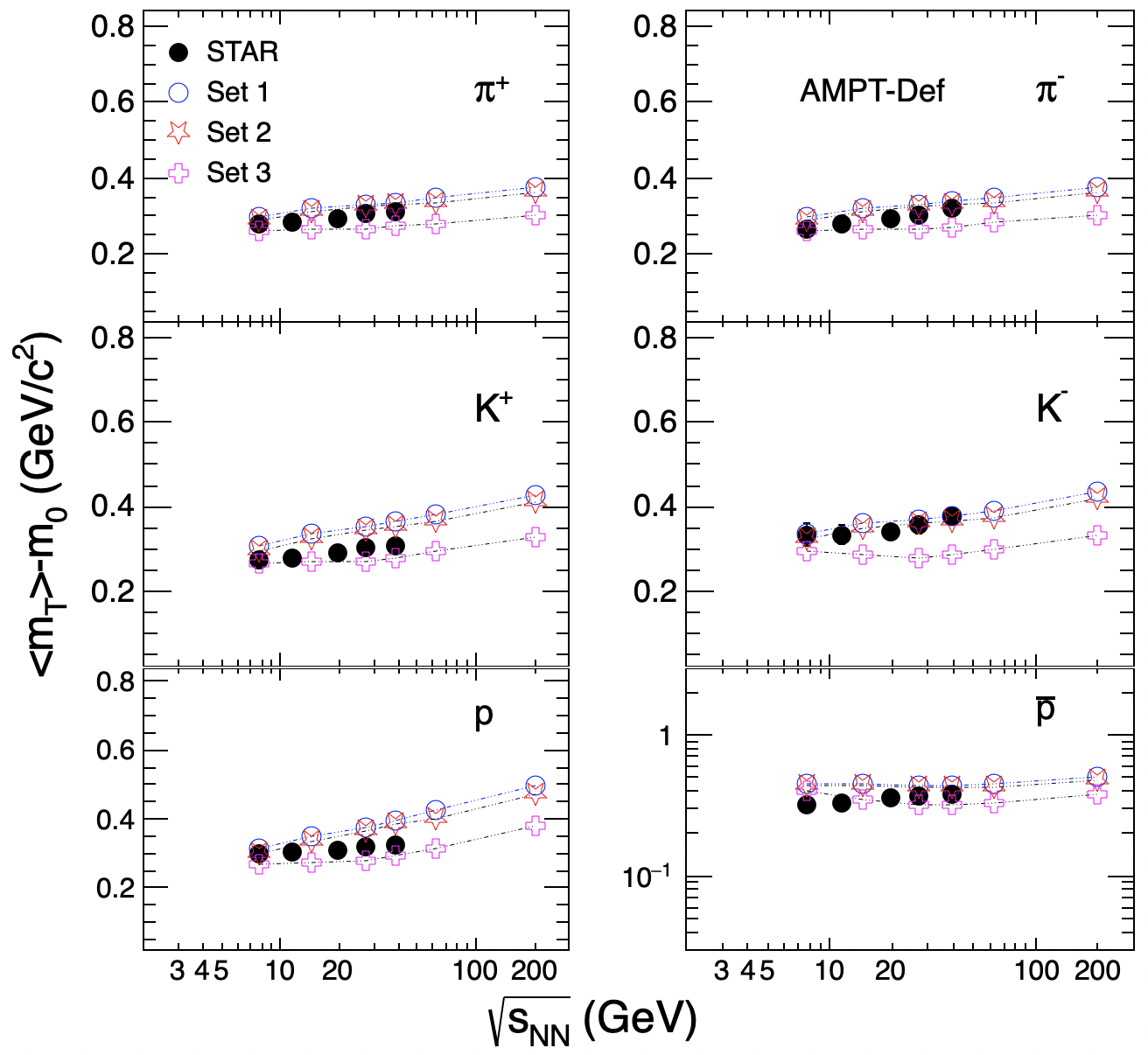}
     \text{(a)}
  \end{minipage}
  \hfill
  \begin{minipage}[b]{0.48\textwidth}
    \includegraphics[height=7.5cm,width=\textwidth]{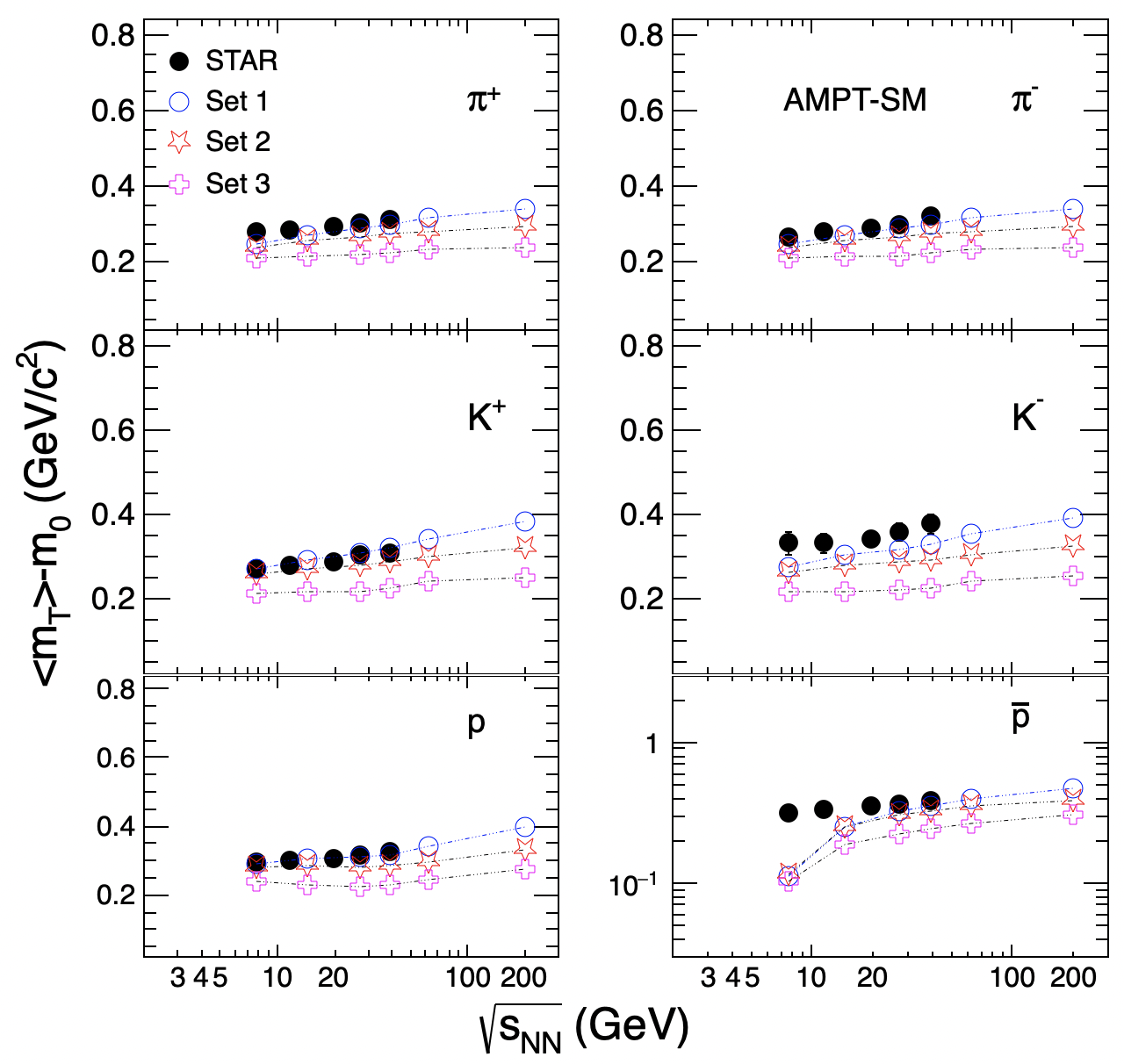}
    \text{(b)}
  \end{minipage}
  \caption{\(\langle m_T \rangle - m_{0}\) of \(\pi^\pm\), \(K^\pm\), \(p\), and \(\bar{p}\) as a function of \(\sqrt{s_{\text{NN}}}\) at midrapidity (\(|y| < 0.1\)). Results are presented for most central Au+Au collisions using AMPT-Def (left panel) and  AMPT-SM (right panel). The AMPT results shown in open markers are compared with the corresponding experimental data (solid marker).}
  \label{fig:your_label}
  \end{figure*}


For a thermodynamic system, average transverse mass, ($\langle m_T \rangle - m_0$), could be an indicative of the temperature of the system, where $m_0$ is the rest mass of the particle. It has been suggested that the energy dependence of $\langle m_T \rangle - m_0$ may be a possible signature of first-order phase tranistion between the hadronic medium and the QGP \cite{VANHOVE1982138}.
Figure 7(a) and 7(b), show the energy dependence of \(\langle m_T \rangle - m_0 \) for $\pi^{\pm}$, $K^{\pm}$, $p$, and $\bar{p}$ in most central Au+Au collisions calculated using AMPT-Def and AMPT-SM, respectively. We observe that all the three sets qualitatively capture the trend of the data.

Figure 8(a) and 8(b) show the energy dependence of 
\(\langle m_T \rangle - m_0 \) for 
$K_s^0$, $\Lambda$, $\bar{\Lambda}$ and $\phi$ in most central Au+Au collisions from AMPT-Def and AMPT-SM, respectively. We observe that Set-1 of AMPT-Def version is describing the data well, however, all the sets implemented in AMPT-SM tend to underpredict the data.


\begin{figure*}[!hbtp]
  \centering
  \begin{minipage}[b]{0.48\textwidth}
    \includegraphics[height=5cm,width=\textwidth]{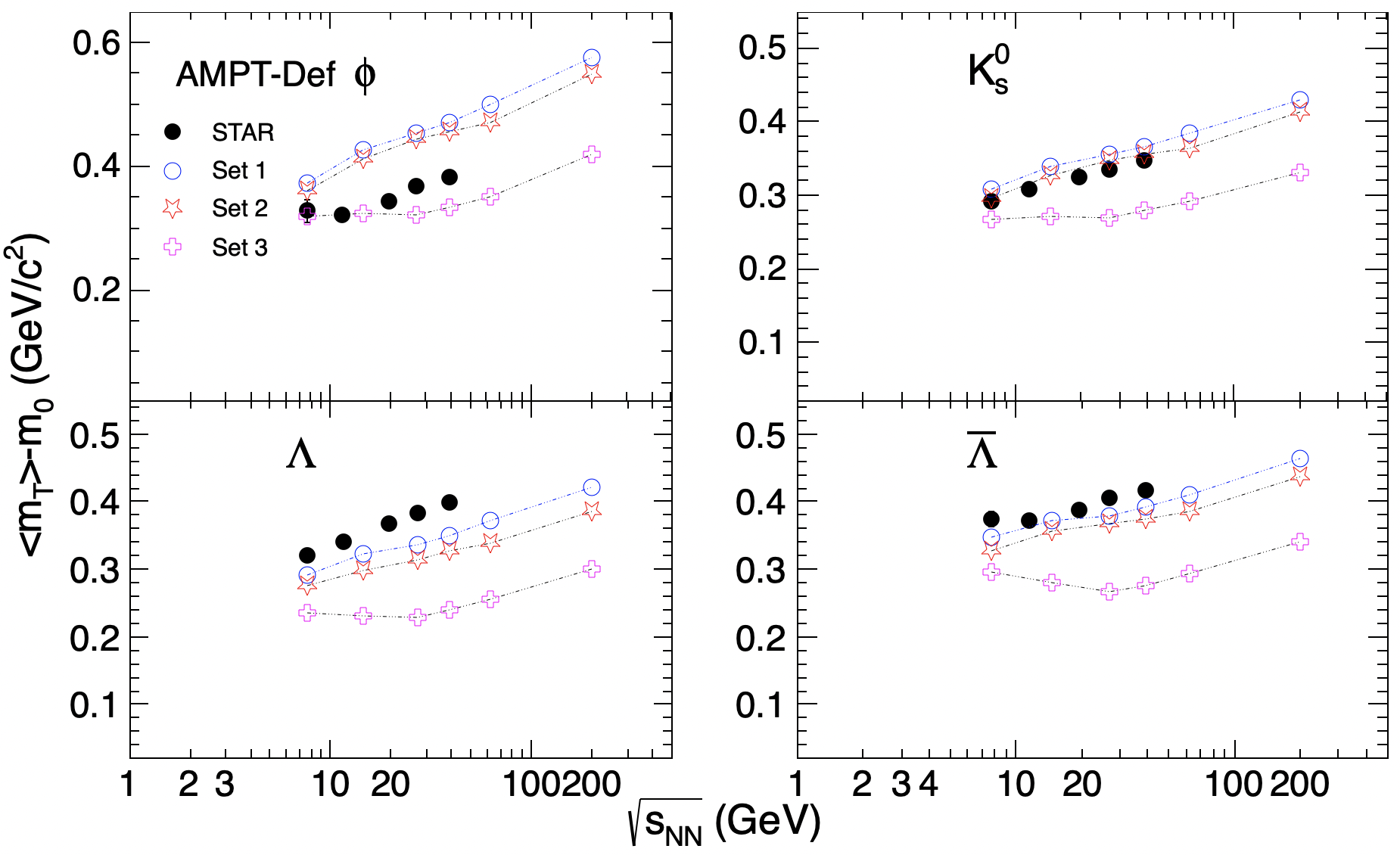}
    \text{(a)}
  \end{minipage}
  \hfill
  \begin{minipage}[b]{0.48\textwidth}
    \includegraphics[height=5cm,width=\textwidth]{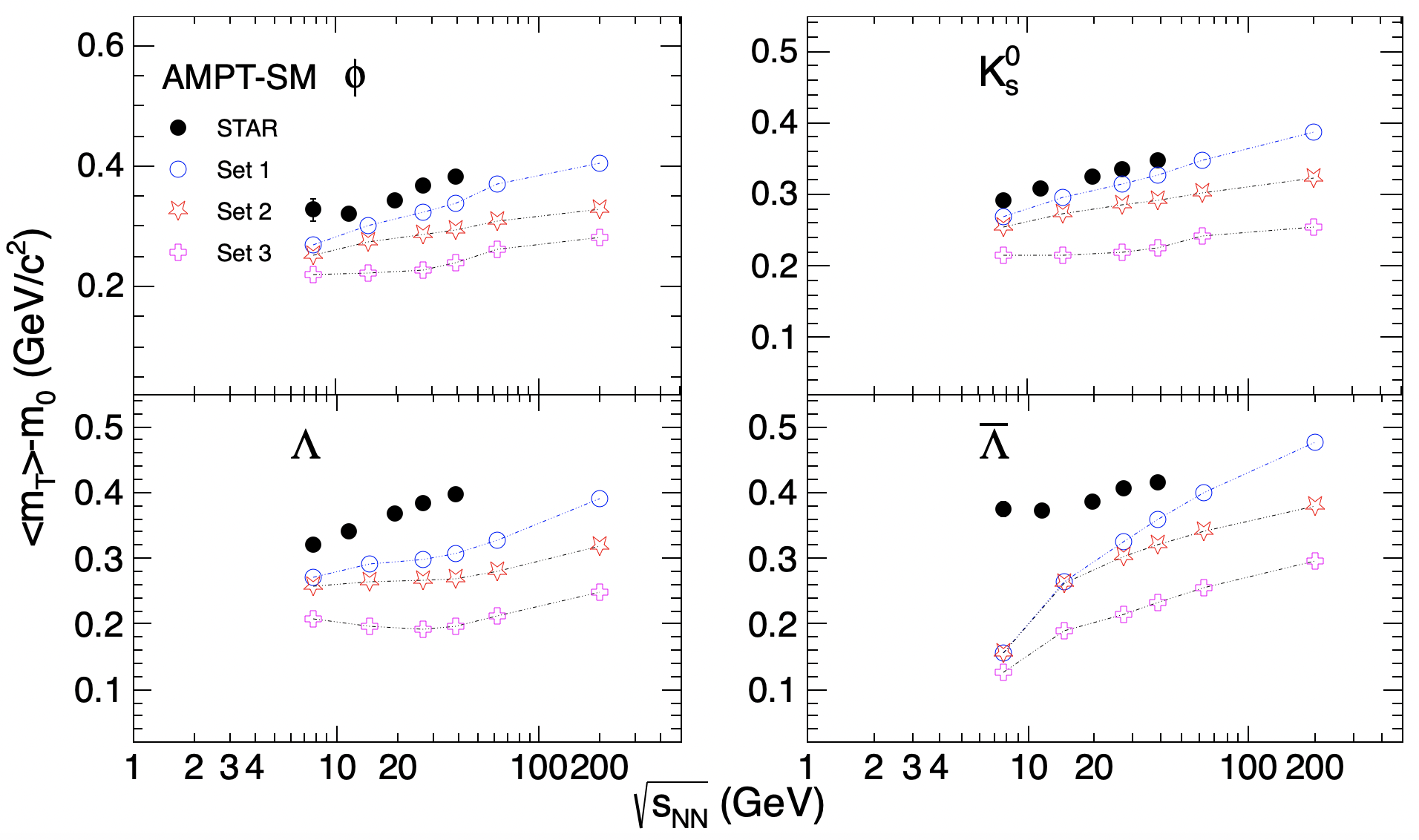}
    \text{(b)}
  \end{minipage}
   \caption{\(\langle m_T \rangle - m_{0}\) of  $K^{0}_{S}$, $\Lambda$, $\phi$, and $\bar{\Lambda}$ as a function of \(\sqrt{s_{\text{NN}}}\) at midrapidity. Results are presented for most central Au+Au collisions using AMPT-Def (left panel) and  AMPT-SM (right panel). The AMPT results shown in open markers are compared with the corresponding experimental data (solid marker).}
  \label{fig:your_label}
  \end{figure*}

The energy dependence of antiparticle-to-particle ratios helps in understanding the particle production mechanism in heavy-ion collisions \cite{STAR:2017sal}. Figure 9 shows the collision energy dependence of the particle ratios \(\pi^-/\pi^+\), \(K^-/K^+\), and \(\bar{p}/p\) in most central Au+Au collisions calculated using the AMPT model. The comparison of the calculations of the AMPT-Def with the experimental data \cite{LACASSE1996153,PhysRevC.57.R466,200053,PhysRevC.55.1420,PhysRevC.60.064901,PhysRevC.60.044904,PhysRevLett.88.102301} are shown in the upper panel, while the lower panel shows the comparison with the AMPT-SM calculations. We observe that the \(\pi^-/\pi^+\) ratio is greater than unity at low energies, due to a significant contributions from resonance decays, such as from $\Delta$ baryons. The \(K^-/K^+\) ratio decreases with decreasing energy due to the associate production of $K^+$. \(\bar{p}/p\) is also observed to decrease with decreasing energy due to the effects of baryon stopping at lower energies. As the energy increases both \(K^-/K^+\) and \(\bar{p}/p\) ratios approaches unity. The systematic effects resulting from variations in model parameters cancel out in particle ratios, leading to an absence of energy dependence on input parameters. We observe that all the sets of both AMPT-Def and AMPT-SM are able to describe the data well.

\begin{figure*}[!hbtp]
  \centering
  \includegraphics[height=8cm,width=0.7\textwidth]{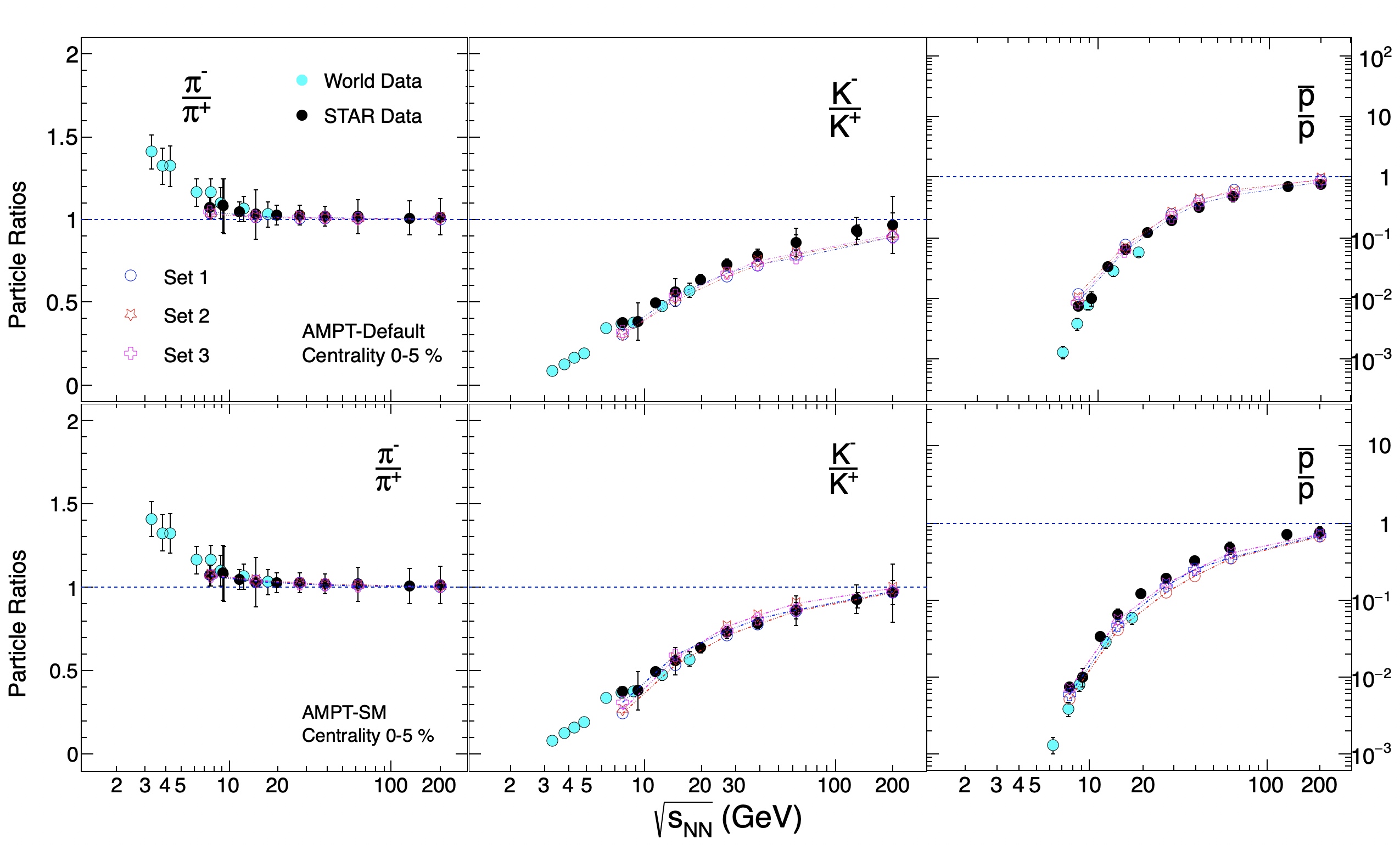}
 \caption{\(\pi^-/\pi^+\),  \(K^-/K^+\), and \(\bar{p}/p\) ratios at midrapidity (\(|y| < 0.1\)) in central most Au+Au collisions at \(\sqrt{s_{NN}} = 7.7, 27, 39, 62.4 \) and \(200\) GeV using AMPT-Def (upper panel) and  AMPT-SM (lower panel).  The AMPT results shown in open markers are compared with the published experimental data (solid marker).}
  \label{fig:your_label}
\end{figure*}

\subsection{\label{sec:C} Kinetic freeze-out}

In this section, we discuss kinetic freeze-out parameters obtained using a blast-wave model in Au+Au collisions at various centre-of-mass energies. In this approach, we perform a simultaneous fit of $p_T$ spectra of $\pi^{\pm}$, $K^{\pm}$, $p$, and $\bar{p}$ with the blast-wave model similar to how it has been done with experimental data \cite{PhysRevC.96.044904} . The blast-wave model is a hydrodynamically motivated model that provides an accurate description of data at low $p_T$. However, it is not well-suited for describing hard processes at high $p_T$. The blast-wave model assumes a common radial flow velocity profile and a thermal freeze-out temperature for all particles \cite{PhysRevC.79.034909}. Assuming a radially boosted thermal source with kinetic freeze-out temperature ($T_{kin}$) and transverse radial flow velocity  ($\beta$),  the  $p_T$ distribution of the particles is given by:

\begin{equation}
\begin{split}
\frac{dN}{p_Tdp_T} &\propto m_T\int_0^R r dr  I_0 \left(\frac{p_T \sinh \rho(r)}{T_{kin}}\right) \\
&\quad \times K_1 \left(\frac{m_T \cosh \rho(r)}{T_{kin}}\right)
\end{split},
\end{equation}

where $m_T = \sqrt{p_T^2 + m_0^2}$ and $\rho(r) = \tanh^{-1}(\beta)$. $I_0$ and  $K_1$ are modified Bessel functions of the first and second kind, respectively. $\beta = \beta_S \left(r/R\right)^n$ is the flow velocity, where $\beta_S$ is the surface velocity, $r/R$ is the relative radial position in the thermal source, and $n$ is the exponent of the flow velocity profile.

\begin{figure*} 
  \centering
  \includegraphics[height=7cm,width=0.7\textwidth]{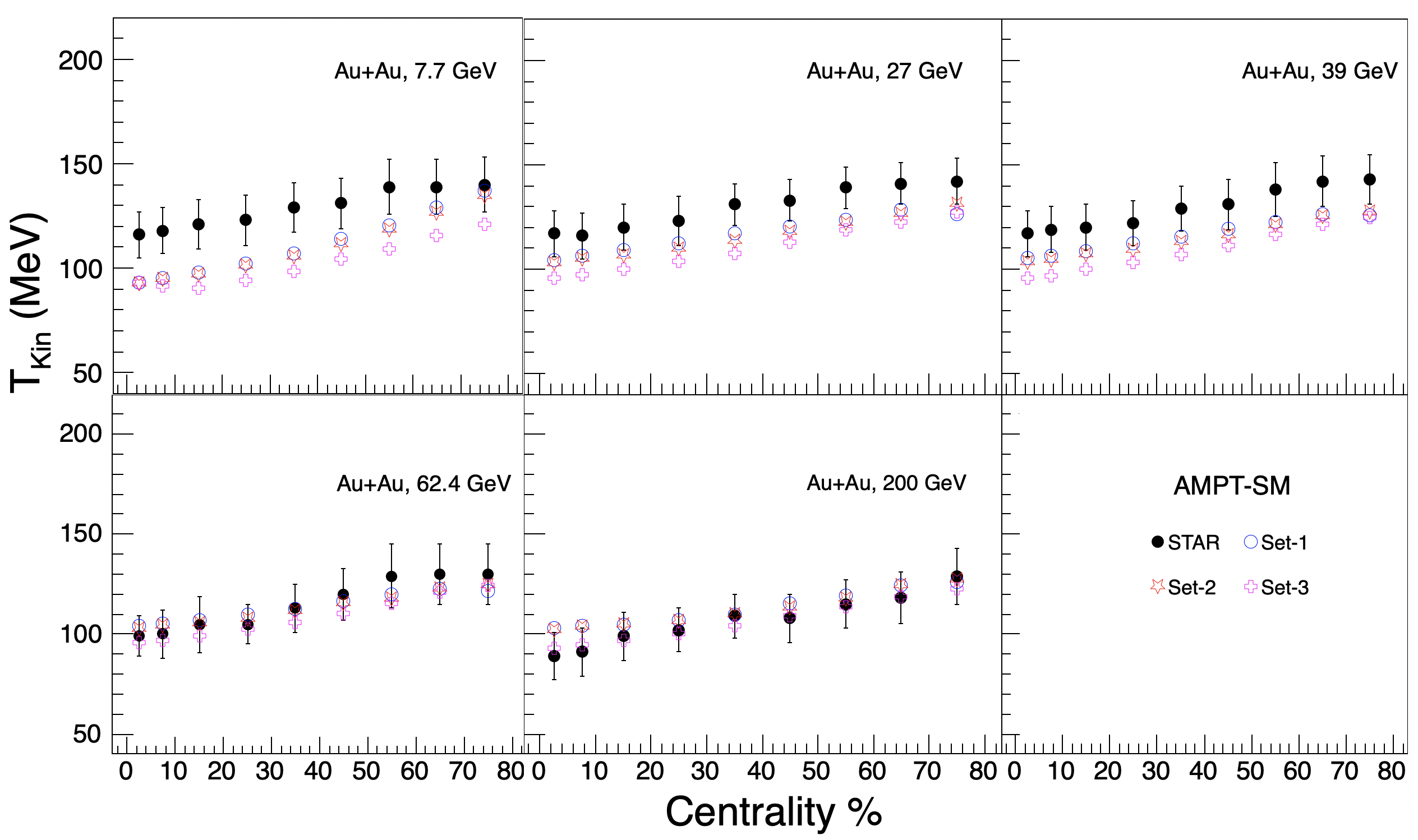}
  \caption{Centality dependence of kinetic freeze-out temperature ($T_{kin}$ in MeV) in different collision energies for four different Sets in Au+Au collisions using AMPT.}
  \label{fig:tkin_centrality}
\end{figure*}

Figure~\ref{fig:tkin_centrality} shows the energy and centrality dependence of $T_{kin}$ in Au+Au collisions at $\sqrt{s_{NN}}$ = 7.7, 27, 39, 62.4, and 200 GeV. We observe that predictions from all the sets of input parameters to the AMPT-SM model are able to capture the decreasing trend of $T_{kin}$ with increasing centrality, consistent with the experimental data. 

\begin{figure*} 
  \centering
  \includegraphics[height=7cm,width=0.7\textwidth]{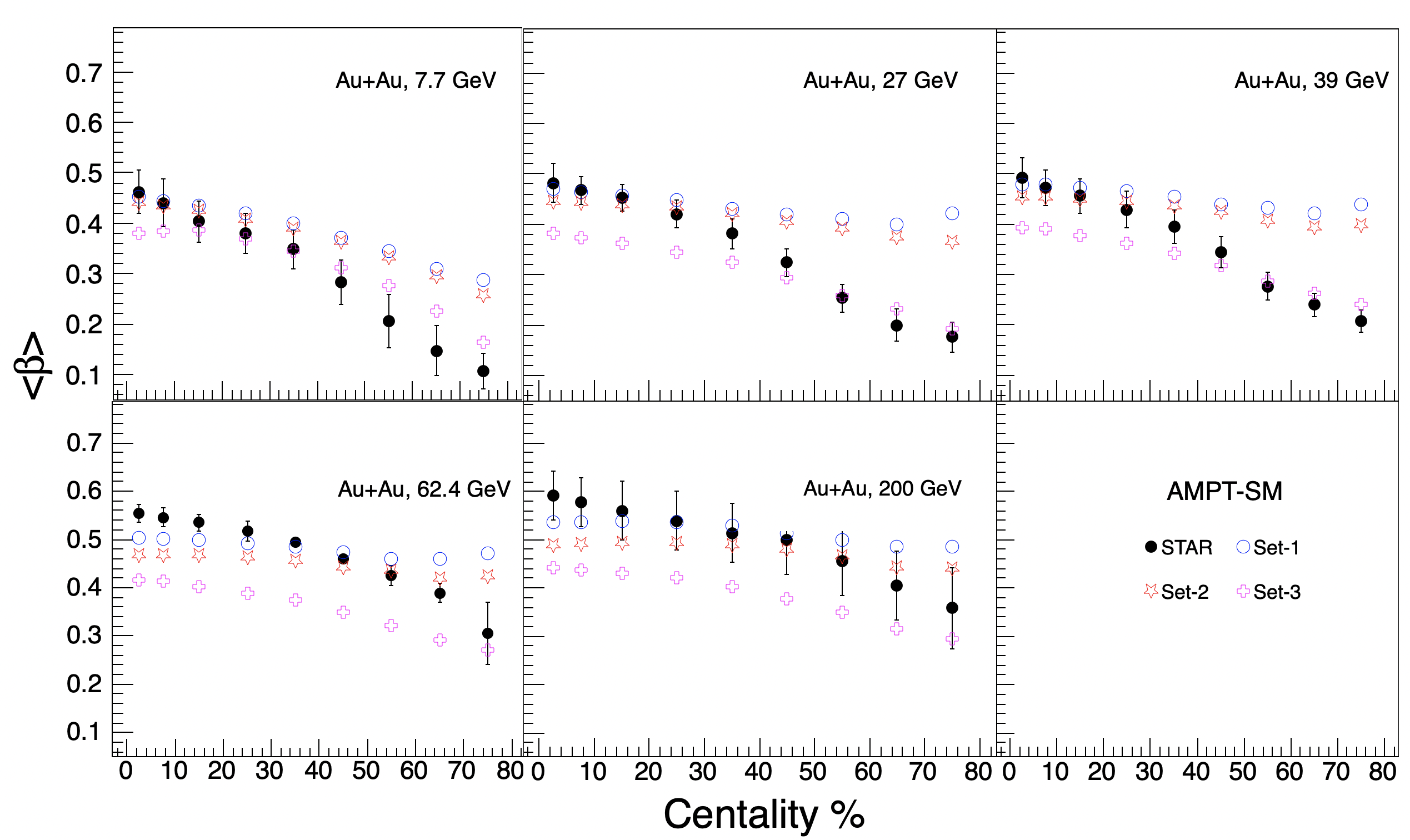}
  \caption{Centality dependence of $\langle \beta \rangle$ in different collision energies for four different Sets in Au+Au collisions using AMPT.}
  \label{fig:beta_centrality}
\end{figure*}

Figure~\ref{fig:beta_centrality} shows the variation of $\langle \beta \rangle$ with centrality of Au+Au collisions at $\sqrt{s_{NN}}$ = 7.7, 27, 39, 62.4, and 200 GeV. We observe that qualitatively all the different sets of input parameters of the AMPT-SM model are able to describe the increasing trend of $\langle \beta \rangle$ with increasing centrality.

\begin{figure*} 
  \centering
  \includegraphics[height=7cm,width=0.7\textwidth]{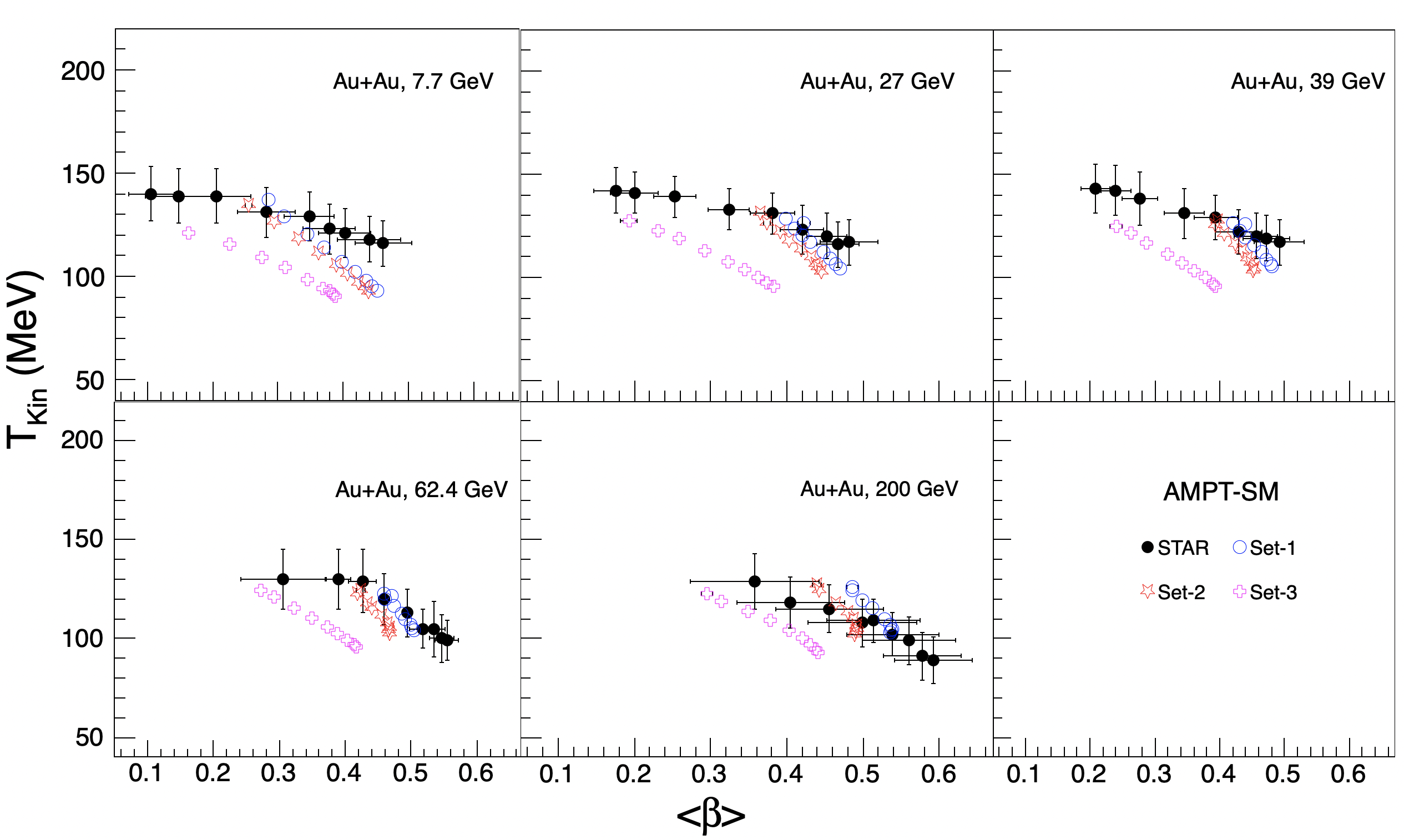}
  \caption{Variation of \(T_{\text{kin}}\) with \(\langle \beta \rangle\) for four different Sets for different collision energies. The centrality increases from left to right for a given energy.}
  \label{fig:tkin_vs_beta}
\end{figure*}

 An anticorrelation plot between $T_{kin}$ and $\langle \beta \rangle$ is shown in Fig.~\ref{fig:tkin_vs_beta}. We observe that all the different configurations of the AMPT-SM model capture this anti-correlation behavior, although Set-3 shows a significant deviation from the data compared to the other two sets.

\section{\label{sec:level4}Summary}
This study aims to understand the particle production mechanism in Au+Au collisions at STAR energies in the framework of the AMPT model covering a wide range of $\sqrt{s_{NN}}$ = 7.7, 27, 39, 62.4, and 200 GeV. We have studied the production of $\pi^{\pm}$, $K^{\pm}$, $p$, $\overline{p}$, $K^{0}_{S}$, $\Lambda$, $\bar{\Lambda}$, and $\phi$ in most central Au+Au collisions at mid-rapidity. In addition, various bulk properties including the transverse momentum spectra, particle yield, the mean transverse mass, and particle ratios, as well as freeze-out properties have been studied. Both the Default and String Melting versions of the AMPT model initialized with different sets of Lund string fragmentation parameters (as listed in Table~\ref{table:5x5}) are compared with data from the STAR experiment. 

We observe that the $p_T$ spectra of identified hadrons are well-described by Set-2 of AMPT-SM at higher energies and by Set-2 of AMPT-Def at lower energies. For strange hadrons we conclude that Set-1 of AMPT-Def version is describing the data well at the studied collision energies. The comparison of bulk properties with the results from the AMPT model suggests that: 

\begin{itemize}
    \item Set-2 of input parameters describe the $dN/dy$ of $\pi^{\pm}$, $K^{\pm}$, $p$, and $\overline{p}$ reasonably well for both AMPT-Def and AMPT-SM.
    \item Set-1 and Set-2 of input parameters describe the $dN/dy$ of $K^{0}_{S}$, $\Lambda$, $\bar{\Lambda}$, and $\phi$ reasonably well.
    \item We observe that all three sets qualitatively capture the trend of \(\langle m_T \rangle - m_0\) for all studied particle as observed in the data.

    \item All the three sets of input parameters are able to describe the particle ratios reasonably well for both the versions of the AMPT model.

    \item All three sets of input parameters of AMPT-SM qualitatively capture the decreasing trend of \(T_{kin}\) and the increasing trend of \(\langle \beta \rangle\) with increasing centrality. Additionally, they capture the anti-correlation between \(T_{kin}\) and \(\langle \beta \rangle\), as observed in the experimental data.

\end{itemize}

In conclusion, we observe that the bulk properties studied in the framework of the AMPT model are sensitive to the input parameters. Comparing the AMPT results with the experimental data helps in understanding the underlying particle production mechanism. The results presented in this paper may help in deciding a better set of input parameters for future studies. 


\begin{acknowledgments}
CJ acknowledges the financial support from DAE-DST Project No. 3015/I/2021/Gen/RD-I/13283 and KN is supported by OSHEC, Department of Higher. Education, Government of Odisha, Index No. 23EM/PH/124 under MRIP-2023. 
\end{acknowledgments}

\bibliography{apssamp}

\end{document}